\documentclass{emulateapj}
\def\lsim{\lower.5ex\hbox{$\; \buildrel < \over \sim \;$}}
\def\gsim{\lower.5ex\hbox{$\; \buildrel > \over \sim \;$}}
\def\abeq{\lower.7ex\hbox{$\; \buildrel \sim \over - \;$}}

\def\t{\ifmmode {\tau} \else $\tau$ \fi}

\def\ref{\noindent \hangafter=1 \hangindent=0.7 truecm}

\def\cm{\ifmmode {\rm cm}^{-1} \else cm$^{-1}$ \fi}
\def\s{\ifmmode {\rm s}^{-1} \else s$^{-1}$ \fi}
\def\cc{\ifmmode {\rm cm}^{-3} \else cm$^{-3}$ \fi}
\def\cs{\ifmmode {\rm cm}^{-2} \else cm$^{-2}$ \fi}
\def\g{\ifmmode \gamma \else $\gamma$\fi}
\def\G{\ifmmode \Gamma \else $\Gamma$\fi}

\def\kms{\ifmmode {\rm km\ s}^{-1} \else km s$^{-1}$\fi}

\usepackage{graphicx}

\begin{document}

\title{The Mid-Infrared Emission of M87}

\author{Eric S. Perlman\altaffilmark{1,2}, R. E. Mason\altaffilmark{3}, Christopher Packham \altaffilmark{4}, N. A. Levenson\altaffilmark{5}, Moshe Elitzur\altaffilmark{5},
Justin J. Schaefer\altaffilmark{4}, Masatoshi Imanishi\altaffilmark{6},
William B. Sparks\altaffilmark{7}, James Radomski\altaffilmark{8}} 
\altaffiltext{1}{Joint Center for Astrophysics, Physics Department, University of Maryland, Baltimore County, 1000 Hilltop Circle, Baltimore, MD 21250}

\altaffiltext{2}{Current Address:  Department of Physics \& Space Sciences,
Florida Institute of Technology, 150 W. University Boulevard, Melbourne, FL  32901}

\altaffiltext{3}{Gemini Observatory, Northern Operations Center, 670 N. A'Ohoku Place, Hilo, HI 96720, USA}

\altaffiltext{4}{Department of Astronomy, University of Florida, 211 BRSC,
Gainesville, FL  32611, USA}

\altaffiltext{5}{Department of Physics and Astronomy, University of Kentucky, 177 Chem.-Phys. Building,  Lexington, KY 40506-0055, USA}

\altaffiltext{6}{National Astronomical Observatory, Mitaka, Tokyo 181-8588, Japan}

\altaffiltext{7}{Space Telescope Science Institute, 3700 San Martin Drive, Baltimore,MD 21218, USA}

\altaffiltext{8}{Gemini Observatory, Southern Operations Center, c/o AURA, Casilla 603, La Serena, Chile} 

\begin{abstract}

We discuss {\it Subaru} and {\em Spitzer Space Telescope}  imaging 
and spectroscopy of M87 in the mid-infrared from 5-35 $\mu$m. 
These observations allow us to investigate mid-IR emission 
mechanisms in the core of M87 and to establish that 
the flaring, variable jet component HST-1 is not a major contributor 
to the mid-IR flux. The {\it Spitzer} data include a
high signal-to-noise 15-35 $\mu$m spectrum of the knot A/B complex in
the jet, which is
consistent with synchrotron emission.  However, a synchrotron model
cannot account for the observed {\it nuclear} spectrum, even when
contributions from the jet, necessary due to the
degrading of resolution with wavelength, are included.  The {\it
Spitzer} data show a clear excess in the spectrum of the nucleus at
wavelengths longer than 25 $\mu$m, which we model as thermal emission
from cool dust at a characteristic temperature of $55 \pm 10$ K, with
an IR luminosity $\sim 10^{39} {\rm ~erg ~s^{-1}}$.  Given {\it
Spitzer'}s few-arcsecond angular resolution, the dust seen in the
nuclear spectrum could be located anywhere within $\sim 5''$ (390 pc)
of the nucleus. In any case, the ratio of AGN thermal to bolometric
luminosity indicates that M87 does not contain the IR-bright torus that
classical unified AGN schemes invoke.  However, this result is consistent with theoretical
predictions for low-luminosity AGNs.  

\end{abstract}

\maketitle

\section{Introduction}

M87, the dominant galaxy in the Virgo cluster, is one of the nearest
(distance=16 Mpc, 1\arcsec = 78 pc) radio galaxies.  Since 1918, when
Heber Curtis observed a ``curious straight ray'' extending from its
nucleus, it has been known to host a bright jet that is visible at
radio through X-ray wavelengths.  The jet, one of the primary
hallmarks of M87's nuclear activity, has a complex, knotty structure
(see Perlman et al. 2001b [hereafter P01b] and references therein).
M87 also exhibits energetic line emission from a disk that is seen to
be in Keplerian motion (Ford et al. 1994, Harms et al. 1994).
Longslit spectroscopy of the material within the disk has allowed the
mass in the inner 3.5 pc of M87 to be measured at $\approx 3 \times
10^9 M_\odot$ (Marconi et al. 1997), implying the presence of one of
the most massive known black holes.

M87's proximity allows us to obtain particularly high spatial
resolution.  Unfortunately, M87 is quite faint in the infrared, with a
core flux of only 16 mJy at 10.8$\mu$m (Perlman et al. 2001a,
hereafter P01a). This makes {\it Spitzer} well-suited to sensitive
observations of the infrared emission from this object, while at the
same time the small spatial scales that can be probed make M87 an
appropriate mid-infrared (mid-IR) target for the largest ground-based
telescopes -- albeit one requiring long integrations.  The knots in
the jet are significantly fainter than the nucleus. Table 1 reviews
the locations and extents of major knots in the jet, along with N-band
fluxes either directly measured or extrapolated from radio/optical
data.

\begin{deluxetable}{cccc}
\tablecolumns{4}
\tablewidth{0pt}
\tablecaption{Known knots in the jet of M87}
\tablehead{
\colhead{Knot} & \colhead{Dist. from Nuc.$^1$} & \colhead {Approx. Extent$^2$} & \colhead {10.8 $\mu$m Flux$^3$} \\
\colhead{Name} & \colhead{(arcsecs)} & \colhead{(arcsecs)} & \colhead{(mJy)}}

\startdata

HST-1$^4$ & 0.8     &  0.5 $\times$ 0.1  &     $<2.3$  \\
D         & 2.7     &  1.8 $\times$ 0.2  &     $1.2  \pm 0.2$ \\
E$^5$     & 6.0     &  0.9 $\times$ 0.3  &     $0.1$ \\
F         & 8.2     &  1.4 $\times$ 0.4  &     $1.3  \pm 0.2$ \\
I$^5$     & 11.2    &  0.6 $\times$ 0.5  &     $0.25$ \\     
A         & 12.4    &  2.2 $\times$ 1.3  &     $6.4  \pm 0.6$ \\
B         & 14.3    &  2.4 $\times$ 1.3  &     $6.3  \pm 0.6$ \\
C         & 17.4    &  1.9 $\times$ 1.7  &     $4.0  \pm 0.6$ \\
\enddata

\tablenotetext{1}{Location of the flux maximum, rather than centroid.}

\tablenotetext{2}{As measured in optical {\it HST} images.}

\tablenotetext{3}{Except where noted, as given in P01a.}

\tablenotetext{4}{Knot  flux is highly variable; the flux maximum is at the upstream end.  The 2 $\sigma$ flux limit given here is derived
from these {\it Subaru} data.}

\tablenotetext{5}{Not detected in Gemini images (P01a); fluxes
are extrapolated from the radio-optical spectrum (P01b).}

\end{deluxetable}

Previous  ground-based observations showed that most of the
mid-IR emission from the core of M87 can be attributed to synchrotron
radiation from the innermost regions of the jet (P01a, Whysong \&
Antonucci 2004), although a minor contribution from dust at $T\lsim 150$ K
could not be ruled out.  These results provided strong evidence that
the dusty torus of classical unified AGN schemes (Antonucci 1993) 
may be absent in low-luminosity AGNs.  Two recent papers have used Spitzer data
to address the mid-IR emission processes of M87.  Bressan et
al. (2006) used 5-20~$\mu$m IRS spectra to examine the issue of
silicate emission from Virgo Cluster galaxies.  They showed that M87,
like other Virgo ellipticals, exhibits a silicate emission feature at
10 $\mu$m, but that this feature is not spatially resolved by Spitzer.
Shi et al. (2007) used IRAC and MIPS imaging data to
re-examine the origin of M87's mid-IR emission. They found that their
photometry of the nucleus plus jet and lobes can be fit by a
combination of two synchrotron power laws (usually breaking at optical
or higher energies), but that in larger
(1\arcmin) apertures an excess is present over that model. This excess
is ascribed to dust in the host galaxy with a luminosity similar to
that observed in other brightest cluster galaxies.

In this paper we discuss new mid-IR imaging
and spectroscopy of M87 using the {\it Subaru} observatory and {\it
Spitzer Space Telescope}.   We combine data from the
IRAC, MIPS and IRS instruments (the latter covering the entire
5-35$\mu$m range) to obtain a more detailed picture of the emission
processes in the nucleus and jet. In \S 2 we present the
observations and data reduction procedures, while in
\S 3 we discuss M87's line and continuum emission.  We close in \S 4 with a
summary and discussion.

\section{Observations and Data}

\subsection{Ground-Based Observations}

Mid-IR imaging and spectroscopy of M87 were obtained on the nights of
UT 2005 April 27 and 28 using the COMICS camera (Kataza et al. 2000)
attached to the Cassegrain focus of the Subaru Telescope (Iye et
al. 2004).  The $320 \times 240$ SiAs IBC array has a plate scale of
$0.129''$/pixel, delivering a $41 \times 31''$ field of view.  The
detector was read-out in correlated quadruple sampling (CQS) mode
(Sako et al. 2003).  The observing conditions on both nights were not
photometric, and the precipitable water vapor was variable, as high as
3-5 mm on April 27, but reaching 5-7mm on April 28.

For the acquisition images, we used the N11.7 filter (central
wavelength 11.67$\mu$m, bandwidth 1.05$\mu$m, 50\% cut-on/off).  To
remove time-variable sky background, telescope thermal noise and
so-called ``$1/f$'' detector noise, we used the standard chop-nod
technique, with a $10''$ chop throw at a position angle of $-69.5$
degrees (measured north through east), to project the jet along the
detector rows and place the reference beam between minima in the jet
emission.  The chop frequency was 0.45 Hz (standard for Subaru/COMICS
observations) and the total on-source time was 700s.

We used a similar setup for the spectroscopic observations, with
similar chop parameters, but without nodding, as is standard practice with COMICS, 
and the broad N band order blocking filter.
The grating and $0.33''$ slit provided a spectral resolution of $\sim
250$ (0.02$\mu$m/pix), dispersing the entire N band across the
array. The superb design and capabilities of COMICS permit on-slit mid-IR
imaging of the polished slit jaws, simultaneous with the spectroscopic
observations. Thus we were able to ensure that the emission from the
nucleus of M87 passed through the slit and into the spectroscopy arm
of COMICS.  This proved to be crucial as differential refraction
between the telescope's optical autoguider and the MIR science beam
moved the object off the slit in as little as 20 minutes at some
airmasses.  The varying, suboptimal conditions meant that M87 was not
detected in many of the spectroscopy frames, and only those in which
the continuum was visible were stacked and used in the final
analysis. The data presented in this paper represent 2100 s of
on-source time.

The data were reduced using IRAF and in-house developed IDL routines.
The difference for each chopped pair (and for each nod-set for imaging),
was calculated, and the results combined until a single
frame was created.  During the reduction process, chopped pairs
obviously compromised by cirrus, high electronic noise, or other
problems were discarded. 

The PSF was measured from observations of the photometric standard.
The measured full-width at half-maximum (FWHM) of the standard was
$0.55''$, which agrees well with the FWHM of the spectral trace.  Flux
calibration was achieved using HD108985 as a flux standard (Cohen et
al. 1992, 1999; Tokunaga 1984) and interpolating the Cohen models to
the proper bandpass.  Absolute errors in flux calibration were
estimated from the variations in the counts through the course of the
nights.  Given the non-photometric conditions under which the data were
obtained, we estimate an accuracy of at best 15\% for the flux
calibration.

 A 5-pixel extraction aperture was used for the spectroscopy data, and
 spectral flux calibration was with reference to the telluric standard
 star, Vega. No attempt was made to correct for slit losses in the
 spectrum of M87, but we note that the flux density in the calibrated
 spectrum is comparable to that of both the Subaru/COMICS image and
 the Gemini/OSCIR photometry of P01a.


In Figure 1 we show the {\it Subaru} image, while in Figure 2 we show
the {\it Subaru} spectrum.

\begin{figure}
\centerline{\includegraphics*[scale=1.0]{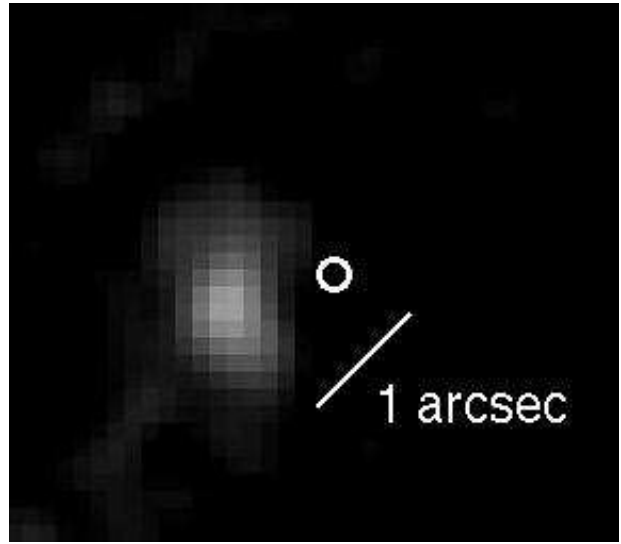}}

\caption{{\it Subaru} acquisition image of M87.  North is up and
east is to the left.  The high spatial resolution of the COMICS
data is apparent, despite its limited sensitivity.  A 3-pixel Gaussian
was used to smooth these data.  We do not detect
emission from knot HST-1, $0.8''$ from the nucleus (location of the circle
on this figure). See
\S\S 2,3 for details. }

\end{figure}

\begin{figure}

\centerline{\includegraphics[scale=0.48]{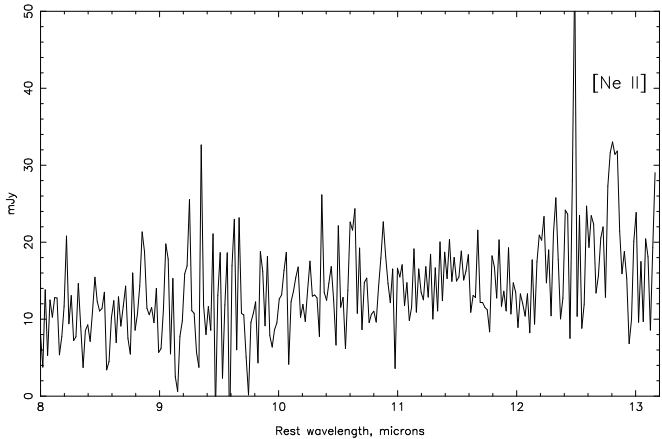}}

\caption{The Subaru/COMICS spectrum of the nucleus, showing the [Ne II] 12.81 
$\mu$m emission line. See \S\S 2,3 for details.}

\end{figure}

\subsection{{\it Spitzer} Observations}

 {\it Spitzer} Space Telescope observations of M87 were obtained from
 the {\em Spitzer} archives. This includes IRS (Infrared Spectrograph,
 Houck et al. 2004) spectra taken on UT 2004 January 4, as well as
 MIPS (Multi-band Imaging Photometer for Spitzer, Rieke et al. 2004)
 images taken on UT 2004 December 26.  Both datasets were acquired as
 part of PID 82 (PI Rieke).  We also obtained data taken on 2005 June
 11 with the IRAC (Infrared Array Camera) as part of PID 224 (PID
 Forman).   The advantage of this particular IRS dataset is that
 it covers the full wavelength range of the short-low and long-low
 modules.  We also comment below on other {\it Spitzer} datasets.
 

The IRAC and MIPS data were processed according to the standard
pipeline recipes.  In both, we used the post-basic calibrated data
(post-BCD).  Fluxes were extracted using circular apertures
appropriate to the diffraction limited resolution at a given
wavelength.  Background subtraction was done using annular apertures
that included all the flux from the galaxy itself.  The {\it Spitzer}
imaging data (shown in Figure 3) clearly show emission from both the
nucleus and the knot A/B complex, as well as extended emission from
the galaxy and the south-west hotspot, a feature thought to be
associated with the unseen counterjet (Hines et al. 1989, Sparks et
al. 1992, Stiavelli et al. 1992).  However, individual knots are not
resolved at {\em Spitzer}'s angular resolution ($\approx 3''$ at $10
\mu$m).  In this paper, we discuss only the emission from the core and
jet knots A and B.  We subtracted from the core flux measured on
the IRAC data any emission that might be due to galactic emission,
using an annulus between $5-15''$; we also subtracted flux due to the
jet by gauging independently the flux in the four quadrants.  
For this reason our fluxes differ somewhat from those
reported in Shi et al. (2007).  The 24$\mu$m
flux derived in this way agrees fully (within 1$\sigma$) with that found
in the IRS spectrum of the core (\S 3.1).

\begin{figure}

\centerline{\includegraphics[scale=0.87]{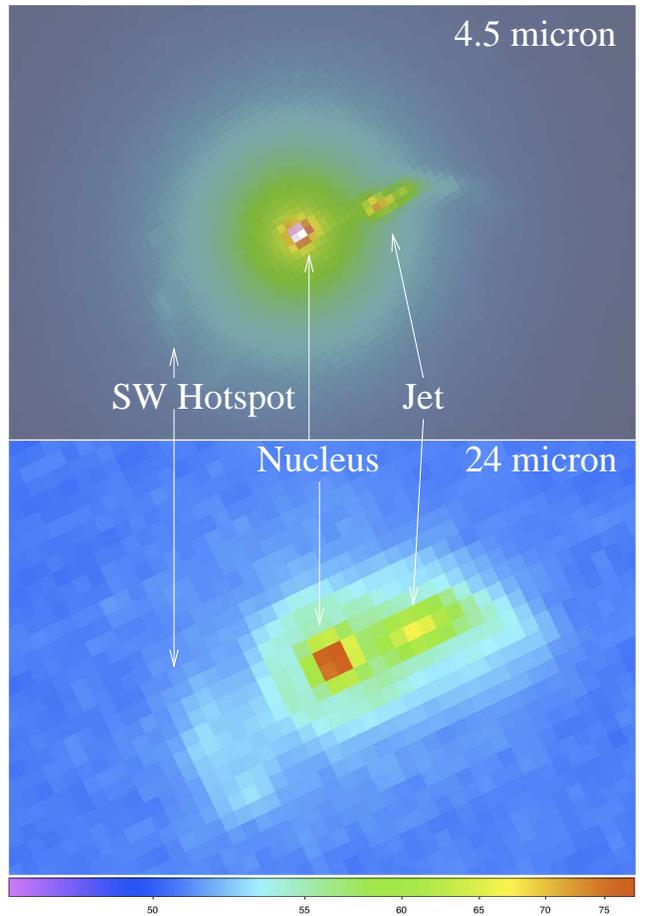}}

\caption{{\it Spitzer} imaging of M87, at 4.5 and 24.5 $\mu$m.  We have
indicated the core, jet and SW hotspot for reference.  See \S 2 for
details.}

\end{figure}



The IRS spectra were acquired using both orders of the low-resolution
short-wavelength (``short-low'') and low-resolution long-wavelength
(``long-low'') modules of the IRS.  Total integration times of 28 s
and 60 s were used for the short- and long-wavelength modules,
respectively, and spectra were obtained in two nod positions separated
by 1/3 of the slit length, translating into nod throws of $19''$ and
$56''$ for the short-low and long-low modules respectively.  The
spacecraft roll angle and relative slit position angles were such that
the $3.6 - 3.7''$-wide short-low slits were oriented perpendicular to
the jet, while the wider ($10.5 - 10.7''$) long-low slits were aligned
along the jet axis and therefore contain emission from the brightest
region of the jet (primarily knots A and B, see Table 1).

After standard pipeline processing, thermal background emission was
removed by subtracting spectra taken at the two nod positions, and 1D
spectra then extracted using the SPICE package. The default extraction
apertures (7.2\arcsec\ at 6 $\mu$m, 14.4\arcsec at 12~$\mu$m, scaling
linearly with wavelength) were employed for the extraction of the
short-low data, whereas to better separate the nuclear and jet
emission in the long-low data, we used a smaller aperture that varied
in size with wavelength, increasing linearly from 5.1'' at 17$\mu$m to
10.2'' at 34 $\mu$m, comparable to the FWHM of {\it Spitzer}'s PSF at
those wavelengths.  As the long-low slits covered both the nucleus and
the knot A/B region, we extracted two 1D spectral datasets.  No
significant fringing is visible in any of the spectra, and no
defringing was attempted.

As M87 is an extended source, it is non-trivial to equalize the flux
calibration of the orders in both the short-low and long-low modules,
or between the two modules, due to the different extraction apertures.
As a result, the spectra from different orders may contain differing
contributions from the various extended structures, e.g., the various
jet components.  
The nucleus is essentially unaffected by this problem at short wavelengths;
however, at longer wavelengths it does have a significant effect.  The fact
that the nucleus is essentially unresolved at short wavelengths 
allowed us to use the flux scale from the pipeline for the SL data.  We then
made use of the overlap and ``bonus segment'' data
to match up the flux and slope of the spectral segments blueward and redward
of each join.  No correction was found to be
necessary to scale the SL1 data to that from SL2. 
However, a scaling factor of 1.67 $\pm 0.03$ was found to
be necessary to scale the LL2 data to the SL1 data, while a scaling
factor of $1.38 \pm 0.03$ was necessary to scale the two long-low
orders of the nuclear spectrum.   

The knot A/B region, by contrast, is highly extended (about $1.3\arcsec 
\times 5\arcsec$, 
as measured on the optical and Gemini-N+OSCIR data).  This makes the flux 
calibration process much more complicated, since the SPICE package is not
optimized (or intended) to deal with extended sources of this nature.
In this case we used the 
Gemini-N + OSCIR result (P01a) to set the flux scale at 10.8$\mu$ m, and then 
use the best-fit spectral index to
extrapolate  from the LL spectra down to that wavelength (note that we do 
not have SL data for the knot A/B region).  We then
used the overlap and ``bonus segment'' data to match up the flux and 
slope of the spectral segments blueward and redward of the LL1/LL2 join
(similarly to what we did for the nucleus).
Because the structural details of the knot A/B region are rather 
different from the nuclear region, a smaller scaling factor 
($1.04 \pm 0.03$) was required to scale the LL1 data to the LL2 flux scale
once it was fixed using the Gemini-N 10.8$\mu$m flux point.  The
resulting IRS spectra of M87's nucleus and jet are shown in Fig. 4.

\section{Results}

The reduced {\it Subaru} acquisition image is shown in Figure 1.  The
core is clearly detected, with a flux of 20.8 $\pm$ 3.5 mJy, consistent
with the findings of P01b to within the $1\sigma$ errors   The
error in the flux calibration is dominated by the variable weather
conditions.  We do not detect any jet features.  Even for the knot A/B
complex this is not unexpected given the extended nature of the knot,
our short integration time, fairly narrow band and the poor sky
conditions.  Surprisingly, however, we do not detect the flaring jet
component HST-1 (Harris et al. 2003, 2005; Perlman et al. 2003),
$0.8''$ from the core, despite the fact that at its peak in 2005 March
(just a few weeks before the {\em Subaru} observations), its
0.8~$\mu$m flux was brighter than that of the core itself (Biretta et
al., in prep.).  We place a limit of 2.3 mJy at $2 \sigma$ on the flux
from knot HST-1.

The COMICS nuclear spectrum is shown in Figure 2.  The signal-to-noise
ratio of the spectrum is low, but we detect continuum emission plus a
single emission line at 12.81~$\mu$m (rest frame).  This spectrum
contains only flux from the nucleus; the extraction aperture would
have excluded any significant contribution from HST-1 which in any
case is faint.  In Figure 4, we show the {\it Spitzer} spectrum of the
knot A/B region (blue) as well as the nucleus (black).  The
signal-to-noise ratio of the {\it Spitzer} data, while still modest
($\sim 10$ per pixel) is significantly better than that of the {\it
Subaru} spectrum.  We have overplotted on Figure 4 a binned version of
the {\it Subaru} nuclear spectrum, which matches well with that
obtained by {\it Spitzer}.  Despite the difference in angular
resolution between the {\it Spitzer} ($\sim 3''$ at 10 $\mu$m) and
{\it Subaru} ($\sim 0.3''$) data, the nuclear spectra are similar in
flux density and spectral slope in both datasets.  This confirms that
HST-1 (which in the {\it Spitzer} data would not be resolved from the
nucleus) makes no more than a minimal contribution to the nuclear
spectrum. Both the noise in the {\it Subaru} data, however, and the
known optical variability of the nucleus and HST-1 (measured to be as
large as 50\% and 500\%, respectively; Tsvetanov et al. 1998, Perlman
et al. 2003) prevent us from setting stricter limits on the mid-IR
flux of HST-1.

\begin{figure}

\centerline{\includegraphics[scale=0.52]{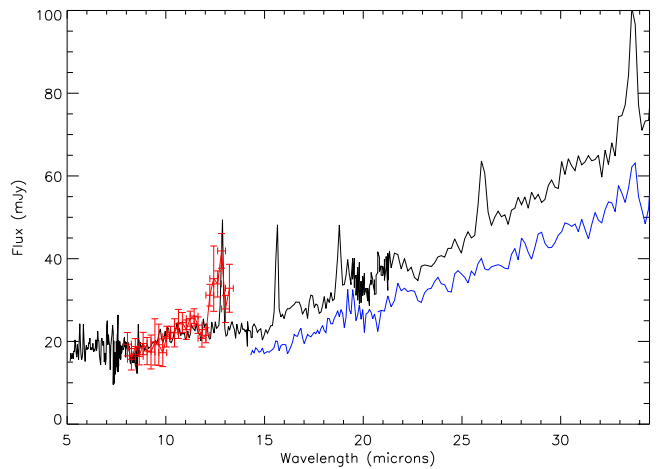}}

\caption{The Spitzer IRS spectrum of the nucleus of M87 (black) as
well as knots A and B (blue).  For comparison, the binned Subaru
COMICS data are shown in red (error bars reflect the standard error on 
the mean of the points in each bin).  To better separate the two IRS
spectra, the data for knots A and B have been multipled by 0.75.
Several low-ionization emission lines
can be seen in the spectrum of the nucleus, 
as well as hints of silicate emission around 10 and 18~$\mu$m.  
See \S\S 3, 3.1 for details.}

\end{figure}

The overall similarity of the spectral shapes of the nucleus and the
knot A/B complex, at least below $25~\mu$m, confirms the synchrotron
nature of the majority of the nuclear emission, first shown by P01a.
A close examination of the nuclear and jet spectra does, however,
reveal differences.  The first of these is the presence of five
obvious emission lines in the spectrum of the nucleus, whereas only
one weak emission line is seen in the knot A/B spectrum.  A more
subtle difference is that the nuclear spectrum contains a sharper
upturn towards longer wavelengths.  We discuss the spectral data in
more depth in the following subsections.  Note that in the discussion
that follows, we use the terms ``nuclear region'' and ``nuclear
spectrum'' to refer to the {\it Spitzer} spectrum taken at the
position of the nucleus (as opposed to that of the knot A/B region),
while what we term as the core refers only to the regions within $\sim
0.3''$ of the central black hole (and thus unresolvable in
ground-based N-band images).

\subsection{IR Continuum Emission}
\label{cont}

The infrared continuum of a galaxy can contain emission from
a number of different components.  In the case of M87, the most likely
sources are thermal emission from warm or cool dust, synchrotron
emission from the jet, and (at shorter wavelengths) starlight from K
and M stars.  In order to determine the nature and origin of this IR
emission, it is necessary to correctly account for all possible
sources that could fall within the slit and beam of {\it Spitzer} at
any given wavelength.  At Spitzer's angular resolution the nuclear
spectrum will contain a contribution from emission in the jet (and
vice versa; see Table 1), and, because of the decrease in spatial
resolution towards longer wavelengths, the magnitude of that
contribution will increase with wavelength. We have modeled both the
nuclear and knot A/B spectra, and accounted for this effect as
follows.

We model each component as a 1-dimensional Gaussian of width equal to
Spitzer's diffraction limit (which we take to be a Gaussian of FWHM
$2.96''$ at 10~$\mu$m, located at its flux maximum position (Table
1)).  We then integrate each Gaussian across the slit (see \S 2.2 for
details).  We use the fluxes of jet knots C, D and F of P01a and take
their positions and radio-optical spectral indices $\alpha_{\rm RO}$
from P01b [$F_\nu \propto \nu^{-\alpha}$]. For fainter knots, we take
2.1 $\mu$m fluxes and positions from P01b and extrapolate to 10$\mu$m
using the $\alpha_{\rm RO}$ values in P01b.  For knots A and B, we fix
the fluxes to be equal to those observed by P01a but allow their
spectral index to vary.
We do not include a contribution from HST-1 given that we do not
detect it in our 11.7 $\mu$m image (but see below). 
The modeling of the knot A/B region spectrum is shown in Figure 5.  We
show separate curves for the contributions of the core, knot A/B and
other knots to the observed spectrum.  This modeling procedure can
satisfactorily account for the observed spectrum of the knot A/B
region.  We fit a power-law to the spectrum of the knot A/B emission
by holding constant the contributions from all other components to the
fluxes and spectral indices from P01a and P01b, with the exception of
the core (for which we used the power-law fit we detail below).  Once
this is done, we obtain a best fit spectral index for knots A and B
of $\alpha_{\rm IR}=0.75 \pm 0.04$.  This is within $1\sigma$ of the
radio-optical spectral indices given in P01b for these knots.  Thus
the spectrum of the knot A/B region is entirely consistent with
synchrotron emission, as previously modeled by P01a, P01B and Shi et 
al. (2007).

\begin{figure}

\centerline{\includegraphics[scale=0.52]{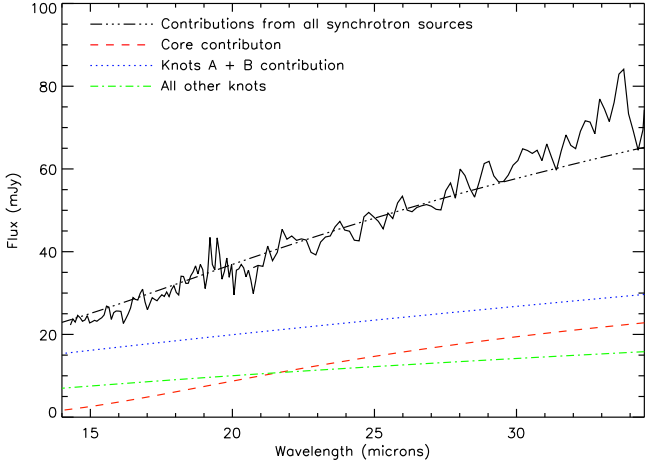}}

\caption{Spitzer IRS spectrum of jet knots A and B, modeled as discussed 
in \S 3.2.  In blue, we show the contribution due to knots A
and B (summed); in red, the contribution from the core, and in green
the contribution from all other knots.  The black line represents the
sum of the contributions from the core and all knots.  The observed
spectrum for knots A and B can be satisfactorily reproduced with these
synchrotron components.}

\end{figure}

In Figures 6 and 7, we show the modeling of the nuclear spectrum, with
Figure 6 showing the $5-35 ~\mu$m region alone and Figure 7 extending
it to longer wavelengths.  To fit the nuclear spectrum, we need to add
to the jet components an increase of flux of 15\% over that observed
at 10.8$\mu$m by P01a.  Significant nuclear variability in M87 has
previously been noted in the optical by Tsvetanov et al. (1998) and
Perlman et al. (2003).  In addition, our data require a core spectral
index that is quite different from that predicted by P01a,
specifically $\alpha_{\rm IR}=0.41\pm 0.05$ instead of $\alpha \sim
1.0$ (as used also in other papers, including Shi et al. 2007).  This
indicates a synchrotron peak frequency $\gsim 5 \times 10^{13}$ Hz
(and likely $>10^{14}$ Hz), at least 1-2 decades higher in frequency
than that fit by P01a.  The spectral index was fit only over the
spectral range 7.5-15~$\mu$m where no departure is observed from a 
pure power-law form (see below), but the method was otherwise
identical to the one above for the knot A/B spectrum.  The
different spectral index could be the result of variability, as in
blazars (of which M87 is a somewhat misaligned example), where
increases in flux are often accompanied by spectral hardenings and
large increases in peak frequency (see e.g., Pian et al. 1997).  
As already noted, the nucleus of M87 is known to be variable in the 
optical (Tsvetanov et al. 1998, Perlman et al. 2003) on timescales of
$\sim 1$ month, and on even shorter timescales in the gamma-rays 
(Aharonian et al. 2006).

\begin{figure}

\centerline{\includegraphics[scale=0.52]{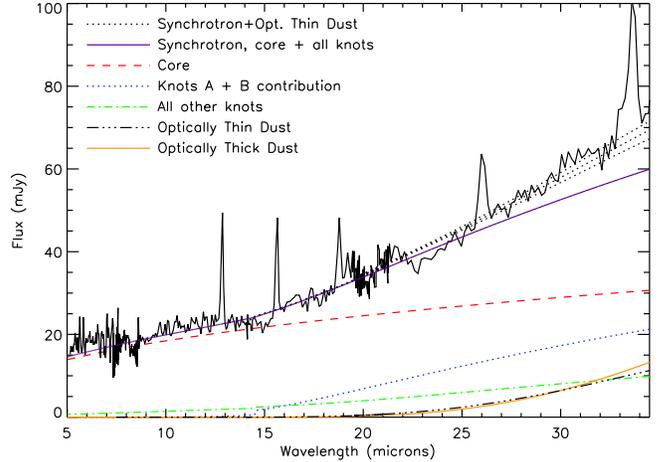}}

\caption{Spitzer IRS spectrum of the nucleus of M87, modeled as discussed 
in \S 3.1.  We follow Figure 5 by showing the
contributions of the core and the knot A/B region as red and blue
respectively, while in green we show the contribution from all other
knots.  This figure shows the sum of these components as purple, and
then optically thin and thick thermal dust emission models as the
lower black and orange curves respectively.  The top set of black
curves represents the sum of all synchrotron emission plus the
optically thin dust model.  The upper and lower black dotted curves represent
the result of multiplyin the 55K thermal model by 1.2 and 0.8 respectively; 
this illustrates the range of normalizations that the data are sensitive to.}

\end{figure}

In contrast to the knot A/B spectrum, we cannot model the observed
nuclear emission with the purely power-law emission characteristic of
synchrotron radiation.  The pure power-law model has discrepancies
both at the short-wavelength and long-wavelength end of the IRS
spectrum. There are two likely causes of the apparent short-wavelength
excess.  The first possibility is that it is due to M and K stars in
the nuclear regions of M87, a component that P01a speculated might
account for the faint (7\% of nuclear flux), extended emission found
in the much deeper OSCIR image.  A second possibility is that the
excess might be the low-frequency tail of the emission from knot
HST-1, which {\it Spitzer} would be unable to resolve from the nucleus
itself.  If we were to use a power-law to extrapolate the observed
excess at $\sim 5
\mu$m down to a wavelength of 0.8~$\mu$m, the flux from this
additional component would equal or slightly exceed that of the core,
consistent with the observed 0.8~$\mu$m flux of knot HST-1 in January
2005 (Biretta et al., in prep.). 

At longer wavelengths, beyond about $23~\mu$m, we see a much more pronounced
excess (Figure 6).  By 35 $\mu$m the pure synchrotron model -- which
includes emission from the core plus all jet knots (i.e., all known
sources of synchrotron emission in M87) -- underestimates the observed
nuclear spectrum by about 20\%.  This excess could be fit by adding an
additional power-law component with spectral index
$\alpha_{excess}\sim 6$.  However, such a component would be
unphysical given known models of synchrotron radiation (e.g., Leahy
1991); and moreover, our model has already accounted for all known
sources of synchrotron emission.  Also, the addition of such a
steep power-law component would dominate over all the other emission
sources by $\sim 50 ~\mu$m, producing a huge excess by 100 $\mu$m,
something which is ruled out by the longer-wavelength MIPS and IRAS
data (Figure 7){\footnote{Note that the ISO 60$\mu$m and 100$\mu$m fluxes
(Haas et al. 2004) are more than a factor two above all other photometry
at similar wavelengths.}} 
Alternatively, we can fit this excess with
thermal emission (Figure 6), which is not unexpected in the nuclear
regions of M87 given the dust lanes found in HST and ground-based
images (e.g., Sparks et al. 1993, Ford et al. 1994).  
We believe that thermal emission from cool dust in the nucleus is the
most physically realistic explanation for the long-wavelength excess
in the few arcsecond-aperture IRS data (distinct from the host galaxy
dust detected at large radii by Shi et al. (2007) in 1\arcmin-aperture
photometry). In addition, the data and modelling presented here allow us 
to to place
constraints on the mass and temperature of dust in the nucleus of M87.

A single blackbody fit to this component yields adequate fits with
temperatures between 45-65K, with higher temperatures being unable to
reproduce the rise at 23-35 $\mu$m, and lower temperatures resulting
in an overproduction of the long-wavelength fluxes from IRAS and {\it
Spitzer}.  We are unable to constrain whether the dust is optically
thin or thick (both curves are shown in Figure 6, for a characteristic
temperature of 55K); however, given that the dust filaments in the
nuclear regions of M87 are also typically marked by H$\alpha$ emission
(Sparks et al. 1993, Ford et al. 2004), and also given the low X-ray 
absorbing column (Perlman \& Wilson 2005) the optically
thin model may be more likely to be correct.


Figure 7 shows only a 55K thermal component, as does Figure 6.
Although temperatures as low as 45K are compatible with the data, and
in fact a single 45K thermal emission component can account for all the
IRS, MIPS and IRAS data, a single dust temperature this low would mean
that the dust in
the innermost $5''$ of the galaxy contributes much more flux than dust
at radii between $5-20''$.  As M87 is extended at mid-IR wavelengths
(e.g., Figure 3), we consider this unlikely. However, sensitive,
higher angular resolution observations at 30-100 $\mu$m would be
required to put tighter constraints on dust temperatures and optical
thicknesses in the nucleus of M87.

\begin{figure}

\centerline{\includegraphics[scale=0.52]{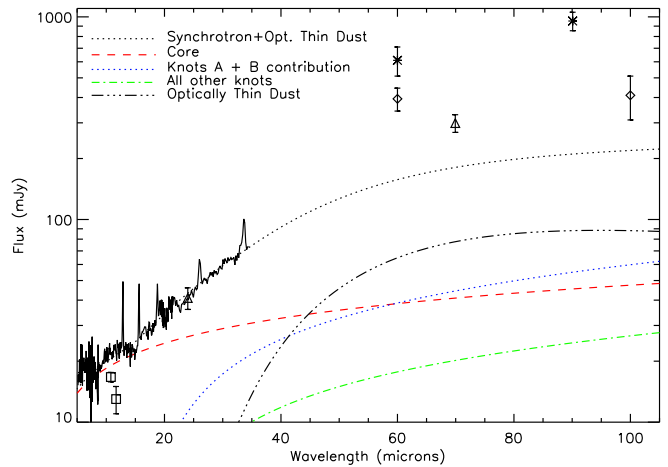}}

\caption{Spitzer IRS spectrum and models of the nucleus of M87, here plotted
with a logarithmic y axis and the wavelength axis continued up to
105$\mu$m. 
All curves are as in Figure 6, with the exception of the optically thin 
thermal emission component, shown in the dash-dot-dot-dot curves (optically
thick thermal emission models are not shown).  Note that longward of
45$\mu$m, thermal emission becomes dominant, a situation that
continues to be the case up to much longer wavelengths.  Individual
points are shown representing data from {\it
Spitzer}/MIPS (triangle), IRAS (diamonds), and ISO (asterisks).
Also shown are (squares) flux points from Perlman et
al. (2001a) and Whysong \& Antonucci (2004), which most likely
lie below our data because of variability.
We do not attempt to fit data at longer
wavelengths; however, if all the dust were at $T<45$K, it would
overpredict the IRAS and {\it Spitzer}/MIPS 70$\mu$m and 160$\mu$m
fluxes. }

\end{figure}

The total luminosity of the thermal component revealed by
the {\em Spitzer} spectra is, for $T=55$K, $9.7 \times 10^{38}$ erg/s
($2.6\times10^5 L_\odot$).  This figure is highly temperature dependent
-- for $T=45$ K we find a thermal luminosity of $3.1 \times 10^{39}$
erg/s, while for $T=65$K, we find a thermal luminosity of $2.9 \times
10^{38}$ erg/s. (By comparison, the error in the normalization of the
thermal component, indicated on Figure 6, is significantly smaller,
approximately 20\%).  Thermal radiation from dust is therefore a
significant contributor to M87's IR emission, being 20-150\% as bright
as the core's synchrotron emission between 1-300~$\mu$m.  It is
interesting to consider how much dust could account for this thermal
emission, and where the dust would be located.  If the dust is heated
entirely by the AGN, it must be within a few parsecs of the core.  If
it is at larger distances it would have to be heated by other
mechanisms, for instance within the star-forming disk
tentatively detected by Tan et al. (2007).
Using standard formulae (Draine 2003 and refs. therein), and assuming 
an optically thin dust component we derive a dust mass of $83 M_\odot$
for a temperature of $T=55$K.  This figure is similarly temperature
dependent: for $T=45$ K we find a dust mass of 180 $M_\odot$, while
for $T=65$ K we find a dust mass of 19 $M_\odot$.  This is a tiny mass
compared to that found in the dusty H$\alpha$ filaments that extend
throughout the galaxy (Sparks et al. 1993). 
It is also tiny compared to the
molecular gas mass derived by Tan et al. (2007) from 230 GHz
observations.

\subsection{Spectral Features}
\label{feat}

The 5-35 $\mu$m spectral region covers a range of dust and molecular
emission and absorption features, as well as many fine structure
lines.  As mentioned above, a line at 12.8 $\mu$m is detected in the
COMICS spectrum of M87.  We ascribe this feature to [NeII] rather than
the 12.7~$\mu$m PAH band because of the lack of other molecular
features (including the stronger 11.2~$\mu$m PAH feature) in both the
COMICS and IRS spectra.  In addition, the IRS spectrum of the nucleus
(Figure 4, 6a) also contains lines of [Ne III] 15.6 $\mu$m, [S III]
18.7 $\mu$m, [O IV]+[Fe II] 25.9 $\mu$m and [S III] 33.5$\mu$m, while
in the knot A/B spectrum (Figures 4, 5), we detect only one emission
line at low significance, 
namely [S III] 33.5$\mu$m. (The [Ar II] 7.0~$\mu$m line in the
spectrum of Bressan et al. 2006 is not clearly detected in our lower
S/N spectrum). Table 2 shows the fluxes of the detected lines in the
nuclear spectrum, obtained by direct integration under each line and
using a continuum level determined by the mean of several points at
either side.


With the exception of [O IV], which may in principle
contribute to the 25.9 $\mu$m
line, the ionization energies of all of the lines are low enough
($<50$ eV) that they can be excited by massive stars and ionizing
shocks. High-excitation lines such as [Ne V] 24.3 $\mu$m, which can be
strong in AGN (e.g. Sturm et al. 2002), are not observed in the
spectrum of M87. Comparison of line fluxes in different apertures
suggest that the line-emitting region may be extended. For instance,
the [Ne II] line in the COMICS spectrum is a factor of about three
weaker than the same line in the larger-aperture Spitzer
spectrum. Although the continuum level for the [Ne II] line in the
COMICS spectrum cannot be determined with great accuracy, such a
pronounced difference implies that the line-emitting material extends
beyond the compact central source observed with COMICS. This
interpretation is supported by the fact that we observe a factor of
about 2.5 lower flux in the [O IV]/[Fe II] line than in the [Fe II]
line in a 14\arcsec\ x 20\arcsec-aperture ISO spectrum of M87 (Sturm
et al. 2002).  The ISO spectrum had sufficient spectral resolution to
separate the [O IV] 25.89 $\mu$m and [Fe II] 25.99 $\mu$m lines but no
lines other than [Fe II] were detected. As expected, all the lines are
unresolved in the $R\sim$100 IRS spectrum.

The lack of high-excitation lines is consistent with the
classification of M87's nucleus as a LINER (a suggestion originally
made by Willner et al. 1985).  This might be considered 
surprising given the luminosity of M87's jet.  However, recent radio
observations suggest that relativistic jets may be quite common among
LINERs (Nagar et al. 2000, Falcke et al. 2000, Filho et al. 2004).
Moreover, due to relativistic beaming,  only a
very small solid angle of the emission-line regions
would be exposed to significant high-energy radiation from the jet.

\begin{deluxetable}{ccc}
\tablecaption{Fluxes of the fine structure lines detected in M87} \tablewidth{0pt}
\tablehead{
\colhead{Line,  $\lambda_{rest}$ \tablenotemark{a}} & \colhead{Flux} & \colhead{E$_{ion}$  \tablenotemark{a}}\\
\colhead{$\mu$m} & \colhead{$10^{-21}\rm ~W ~cm^{-2}$} & \colhead {(eV)}}

\startdata

[Ne II] 12.81  & 1.9 \tablenotemark{b} & 21.6 \\

........       & 5.4 \tablenotemark{c} & 21.6 \\

[Ne III] 15.56 & 4.1 & 41.0 \\

[S III] 18.71 &  2.0 & 23.3 \\

[O IV]/[Fe II] 25.89/25.99 & 1.9 & 54.9/7.9 \\

[S III] 33.48 & 4.5 \tablenotemark{d} & 23.3 \\


\enddata
\tablenotetext{a}{From Sturm et al. (2002)}
\tablenotetext{b}{Subaru/COMICS data}
\tablenotetext{c}{Spitzer/IRS data}
\tablenotetext{d}{See text for discussion}
\label{tab:lines}
\end{deluxetable}

The [S III] 33.5~$\mu$m seen weakly in the knot A/B spectrum (Figures 4, 5)
is unlikely to come from knots A and B, since spectra of M87's jet
have been taken many times in other bands, most recently by HST in the
UV (Gelderman et al. 2005), and have failed to detect line emission,
placing severe limits on the presence of non-stripped nuclei entrained
within the jet.  It is possible that the [SIII] emission originates in
the H$\alpha$ filaments within a few arcseconds of the knot A/B
complex (see Sparks et al. 1993). Alternatively, the emission may
originate from material close to the nucleus, whose light contributes
to the jet spectrum particularly at longer wavelengths where the PSF
is largest. This is consistent with the relative contribution of the
core to the knot A/B spectrum at 33.5$\mu$m, as computed in our
modeling procedure (\S~3.1). That only [SIII] 33.5~$\mu$m, the
longest-wavelength line in the nuclear spectrum, is also observed in
the jet spectrum suggests that this may be the more likely
explanation.

In highly obscured objects, such as Seyfert 2 galaxies, the AGN
unified model predicts absorption features near 10 and 18 $\mu$m due
to the Si-O bond stretch in silicate dust grains. Such features are
indeed observed quite widely in Seyfert 2 galaxies (e.g. Roche et
al. 1991).  Such a feature would not necessarily be expected in M87,
however, due to the fact that its jet (which would be perpendicular to
any dusty torus under unified schemes) is seen at a fairly small angle
to the line of sight ($\lsim 15-20^\circ$, Biretta et al. 1999).
Consideration of the difference between means in points at the peak
and either side of a typical silicate absorption profile in the COMICS
spectrum allows us to put a 95\% confidence limit of $\tau(9.7) \la
0.34$ on any broad silicate absorption feature in the nucleus of
M87. Thus, while a strong absorption feature can be ruled out, this
limit is comparable to the silicate optical depth observed in the
circumnuclear material of Seyfert 2 galaxies such as NGC 1068 ($\sim
0.4$; Roche et al. 1984; Jaffe et al. 2004; Mason et al. 2006).  It is
also consistent with the best available limit ($N_H<6 \times 10^{20}
{\rm ~cm^{-2}}$, much lower than that seen in typical Seyfert 2
galaxies) on nuclear absorption from {\it Chandra} X-ray data (Perlman
\& Wilson 2005), given the recently found correlation between X-ray
N(H) figures and silicate absorption (Shi et al. 2006).

While the COMICS spectrum sets weak limits on the presence of a
9.7~$\mu$m absorption band in the central $<$0.65\arcsec\ of M87, the
IRS spectrum can be used to search for dust features which may exist
on somewhat larger scales. In fact, there are hints of weak silicate
{\em emission} features around 10 and 18 $\mu$m in the IRS spectrum
(the 10$\mu$m feature would be well within the noise of the COMICS
spectrum). The 10 $\mu$m emission feature is also evident in the
IRS spectrum of M87 presented by Bressan et al. (2006; their spectrum
does not cover wavelengths $>20\mu$m).
Previously thought to be rare in AGN, silicate emission features have
now been seen in the LINER NGC 3998 (Sturm et al. 2005) as well as in
several distant quasars (e.g., Hao et al. 2005; Siebenmorgen et
al. 2005).  Although the AGN unified scheme in its most basic form
predicts a silicate emission feature from the hot inner edge of a
dusty circumnuclear torus, the apertures so far used to detect the
silicate feature do not place strong constraints on the location of
the silicate-emitting material in these galaxies. Furthermore, the
cool ($\sim$ 200 K) temperature implied by the strengths and profiles
of the features in the LINER, NGC 3998, suggests an origin in
narrow-line-region dust rather than the inner regions of the torus
(Sturm et al. 2005).  Although Bressan et al. point out that the
10$\mu$m feature in their data is not spatially extended, the PSF of
Spitzer at 10~$\mu$m (3\arcsec) is several times larger than the FWHM
of the nucleus itself in higher-resolution ground-based data.
Therefore it is entirely possible that the silicate emission could be
produced by dust as far as 200 pc from the nucleus.  HST and
ground-based optical observations show the existence of a significant
amount of warm dust in the inner 3'' of M87, as evidenced by the
$H\alpha$ emission in many of these regions (Sparks et al. 1993, Ford 
et al. 1994).  Therefore, the silicate
emission in M87 does not necessarily imply the existence of a nuclear
torus.

\section{Summary and Conclusions}

We have presented mid-IR imaging and spectroscopy of M87, using the
{\it Subaru} observatory and the {\it Spitzer} Space Telescope. The
{\it Spitzer} spectroscopy covered both the nucleus and the knot A/B
complex in the jet. The knot A/B spectrum can be well modeled by
power-law spectra, as expected for synchrotron emission from jet
components, with mid-IR fluxes consistent with those published by P01a
and spectral indices consistent with the optical-radio spectra of
P01b.  However, the {\it Spitzer} spectrum of the nucleus cannot be
modeled by power-law synchrotron emission from the core plus
components in the jet.  We see clear signs of an infrared excess in
the core, which can be well modeled by a thermal spectrum with a
characteristic temperature of $55\pm 10$K.  The low temperature and
luminosity of this component (\S 3.2) are consistent with the limits
found by Perlman et al. (2001a).

Whysong \& Antonucci (2004) made a similar argument based on 11.7
$\mu$m photometry.  The present data support and enhance this result
as the thermal contribution is isolated and measured separately from
the synchrotron contribution.  The measured luminosity is an upper
limit to the amount of emission from a dusty torus in M87; the thermal
contribution measured in the 10.7$\arcsec$-wide IRS slit could arise
anywhere within 420 pc of the nucleus and need not be associated with
a torus at all.
%
%
%
Moreover, given that the bolometric luminosity of  M87 is $\sim 10^{42}
{\rm ~erg ~s^{-1}}$ (e.g., Reynolds et al. 1996), classical torus models
would predict a mid-IR luminosity at least that high (e.g., Risaliti 
\& Elvis 2005).  By comparison we observe a much smaller thermal IR
luminosity ($\sim 10^{39} {\rm ~erg ~s^{-1}}$), making our observations
difficult to reconcile with standard unified models of AGN.

What is the overall significance of the weakness of the torus emission
in M87, both for this galaxy and for FR I radio galaxies as a whole?
One strong possibility is that the weakly-emitting or absent torus in
M87 may be a consequence of the AGN's low luminosity.  In the disk
wind scenario, optically thick regions of an outflowing wind comprise
the geometrically and optically thick ``torus.''  In the model of
Elitzur \& Shlosman (2006), the torus disappears at low luminosities
($\lsim 10^{42} {\rm ~erg ~s^{-1}}$), because accretion onto the
central black hole can no longer sustain the large-scale
outflows. That we do not detect the mid-IR signature of the classical
torus is entirely consistent with the predictions of the disk wind
model.  An alternative to the disk-wind scenario was  
offered by Reynolds et
al. (1996) and di Matteo et al. (2003) who modeled the nucleus of M87
with ADAF models.  Both of those papers derive accretion rates as much
as 4 orders of magnitude below the Eddington rate, which di Matteo et
al. speculate may signal that M87's nuclear activity is at a very late
evolutionary stage.  It is important to note, however, that ADAF models
do not in and of themselves incorporate the notion of a torus.  Finally, 
it is important to point out that the lack of 
significant torus emission in M87 does not yet constitute strong 
evidence that FR Is, as a class, lack luminous tori, although {\it
Spitzer} observations indicate that a significant number may lack 
this component (Ogle et al. 2006, Birkinshaw et al. in prep).  Cen A, for
example, shows significant thermal dust emission in the mid-IR, on
scales ranging from the unresolved sub-pc scale (likely the torus) to
much larger-scale emission (Whysong \& Antonucci 2004, Hardcastle et al. 2006
Radomski et al. 2007).

\begin{acknowledgments}

Based in part on data collected at Subaru Telescope, which is operated
by the National Astronomical Observatory of Japan.  We wish to thank
the Subaru TAC for their support and the Subaru staff, particularly
T. Fujiyoshi, M. Lemmen and M.  Letawsky for supporting our
observations.  ESP acknowledges support from NASA LTSA grants
NAG5-9997 and NNG05-GD63G, as well as HST grant GO-9705.  CP
acknowledges support from NSF grant no. 0206617. NAL acknowledges work
supported by the NSF under grant no. 0237291.  ME acknowledges support
from NSF grant no. 0507421 and NASA grant NNG05-GC38G.  RM and JR were
supported by the Gemini Observatory, which is operated by the
Association of Universities for Research in Astronomy, Inc., on behalf
of the international Gemini partnership of Argentina, Australia,
Brazil, Canada, Chile, the United Kingdom, and the United States of
America. MI is supported by Grants-in-Aid for Scientific Research
(16740117).

\end{acknowledgments}

\vfill\eject

\end{document}